\documentclass{iopconfser}
\usepackage{graphicx}
\usepackage{newtxtext}
\usepackage{newtxmath}
\usepackage{natbib}
\usepackage{hyperref}
\usepackage{enumerate}

\newcommand{\RomanNumeralCaps}[1]

\begin{document}

\title{Improved Analytical Solution for Turbulent Flow in Channel and Circular Pipe}

\author{Alex Fedoseyev}
\affil{Ultra Quantum Inc., Huntsville, Alabama, USA
}
\email{af@ultraquantum.com}

\begin{abstract}
The approximate analytical solution for turbulent flow in a channel was proposed in Fedoseyev (2023). It described the mean turbulent flow velocity as a superposition of parabolic (laminar) and superexponential (turbulent) solutions. The Alexeev Hydrodynamic Equations (AHE), proposed by \cite{Alexeev_1994}, were used as the governing equations to describe turbulent flow. Compared to the Navier-Stokes equations, the AHE include additional terms representing temporal and spatial fluctuations. These additional terms include a timescale multiplier $\tau$, and the AHE reduce to the Navier-Stokes equations in the limit as $\tau \to 0$

In this study, we propose an improved analytical solution formula that provides better agreement with experimental data at high Reynolds numbers. 
The maximum discrepancy between the analytical solution and experimental data has been reduced from 5\% to 2\% for Reynolds numbers of order 100,000, and from 10\% to 4\% for Reynolds numbers up to 35,000,000, based on comparisons with experimental results ranging from the legacy work of Nikuradse (Prandtl group, 1932) to studies by Wei (1989), Zagarola (1996), van Doorne (2007), and the recent work of Pasch (2023).
\end{abstract}


%
%
\section{Introduction\label{sec:intro}}
Experimental investigations of turbulent flow in channels or pipes provide data that can be used to validate numerical simulations, improve theoretical models, and offer new insights into the fundamental mechanisms of turbulence that may not be captured by existing models (Pasch (2023), Van Doorne (2007), Wei (1989), Zagarola (1996)).

The theoretical models fall under Hilbert's   6th problem, Hilbert (1902), which addresses the transition from the
"atomistic view to the laws of motion of continua". 
It involves two steps: (1) from mechanics to kinetics, i.e., from Newton to Boltzmann, and (2) from kinetics to continuum mechanics and nonequilibrium thermodynamics, i.e., from Boltzmann to Euler and Navier-Stokes (Gorban, 2014).
The results of theoretical studies show that the well-known Euler and Navier-Stokes equations are valid only in the "limit of very slow flows with very
small gradients of all fields, i.e., almost no flow at all", Gorban (2014).
Therefore, the search for improved kinetic equations and their corresponding hydrodynamic equations continues. 

The approximate analytical solution proposed for turbulent flow in channel, describing the mean turbulent flow velocity, as a superposition of parabolic and superexponential solutions, Fedoseyev (2023).  The Alexeev Hydrodynamic Equations (AHE) were used as the governing equations for turbulent flow in channel, first provided in  simplified form in \cite{Fedoseyev_2010} as:
\begin{equation}
\it \frac{\partial \bf V} {\partial \rm t} + ({\bf V}\cdot \nabla) {\bf V}
- Re^{-1}\nabla^2 \bf V + \nabla \rm p  = 0,
\label{momeq}
\end{equation}
\begin{equation}
\it \nabla \cdot \bf V = \tau \nabla^2 \rm p ,
\label{newconteq}
\end{equation}
\noindent
where ${\bf V}$ and $p$ are nondimensional velocity and pressure respectively, ${Re=U_0 L_0/\nu}$ - the Reynolds number, $U_0$ - velocity scale, $L_0$ - 
hydrodynamic length scale, $\nu$ - kinematic viscosity, ${\bf F}$ is
 nondimensional body force and nondimensional timescale ${\tau = \tau^* L_0^{-1}U_0}$. Terms with a multiplier $\tau$ are called the fluctuations (temporal and spatial). The complete form of the equations can be found in  \cite{Alexeev_2004}.
 
The AHE include additional terms - representing temporal and spatial fluctuations - compared to the Navier-Stokes equations (NSE). These terms are scaled by a timescale parameter $\tau$, and the AHE reduce to the NSE when $\tau \to 0$. 
The nondimensional $\tau$ is a product of the Reynolds number and the squared length scales ratio, $\tau=Re \cdot (l/L_0)^2$, where $l$ is the apparent Kolmogorov length scale, and $L_0$ is a hydrodynamic length scale. 
The obtained analytical solutions, that depends on two parameters, $Re$ and $\delta$, $\delta= l/L_0=\sqrt{\nu \tau^{*}}/L_0$, agreed rather well with turbulent flow experiments. Here $\nu$ is kinematic viscosity, and $\tau^*$ is material timescale parameter.

The coefficients of superposition in analytical solution were obtained through the minimization principle, the principle of minimum viscous dissipation, Fedoseyev (2024).

In this study, the improved analytical solution formula is proposed, that provides better agreement with experimental data, ranging from Reynolds number $Re=2970$ to $Re=970,000$ from publication starting of legacy paper by Nikuradse (Prandtl group, 1932), to Wei (1989), van Doorne (2007) and research of Pasch (2023).

%
%

\section{\label{sec:solution}Approximate Analytical Solution for Turbulent Flow}

In the analytical solution, the Generalized Hydrodynamic Equations presented by Alexeev (1994) have been employed for turbulent flow in a channel.  
The AHE were derived from the Generalized Boltzmann Equation,
which takes into account finite particle size, Alexeev (1994), while in the traditional Boltzmann equation, particles are treated as material points. The approximate analytical solution, Fedoseyev (2023,) was obtained for a mean turbulent flow velocity $U$ as a superposition
of the laminar (parabolic), $U_L$,
\begin{equation} 
U_L=4y(L-y)/L^{2}
\end{equation}
\noindent and turbulent (superexponential), $U_T$, solutions
\begin{equation}\label{eq:sol_ut}
U_T=1-e^{1-e^{y/\delta}},
\end{equation}
that is
\begin{equation}
U=U_{0}\left[\gamma U_T +(1-\gamma) U_L \right]. 
\end{equation}

The expressions for $U_T$ and $U_L$ are explicitly provided giving the equation

\vspace{-0.25cm}
\begin{equation}\label{eq:u_sol}
U=U_{0}\left[\gamma\left(1-e^{1-e^{y/\delta}}\right)+(1-\gamma)4y(L-y)/L^{2}\right], \end{equation}

\noindent where the coefficients $\gamma$ and $(1-\gamma)$ were introduced,
in 2D channel, $x$ is the coordinate along a channel, $y \leq L/2$ is the transversal coordinate, $L=W/L_0$ is the nondimensional width of a channel with a centerline velocity $U_0$. All parameters are nondimensional. 
The Eq. (\ref{eq:u_sol}) presents a laminar flow velocity if $\gamma = 0$, and turbulent flow velocity if  $\gamma > 0$. 
The parameters $\delta$ is a relative boundary layer thickness scale
\begin{eqnarray}\label{eq:delta}
\delta = {\sqrt{\tau^{*}\nu}}/{L_0},
\end{eqnarray}
\noindent where  $\tau^*$ is the relaxation time, or timescale, a material property for particular liquid or gas  used in the experiments,  $\nu$ is the kinematic viscosity, and $L_0$ is the hydrodynamic scale. The nondimensional $\tau$, a timescale coefficient for the fluctuation terms in AHE, is expressed as
\begin{equation}\label{eq:tau}
\tau = \tau^* L_0^{-1}U_{0}= \delta^2 Re,
\end{equation}
\noindent
where ${Re=U_0 L_0/\nu}$ denotes the Reynolds number. This solution is valid for turbulent flow in circular pipes, if $\delta \ll 1$. An approximate solution for the transverse velocity $V$ was provided in \cite{Fedoseyev_2023} as:
\begin{equation}
V = \frac{1}{\delta Re}(1-e^ {y/\delta}).
\label{eq:v_sol} 
\end{equation}

The solution $U$, Eq.(\ref{eq:u_sol}) agreed well with experimental data for Reynolds numbers up to $Re= 15,000$; however, for larger Reynolds numbers, the deviation becomes noticeable and increases with $Re$.
\begin{figure}[h]
 \begin{center}
 \includegraphics[width=0.60\linewidth]{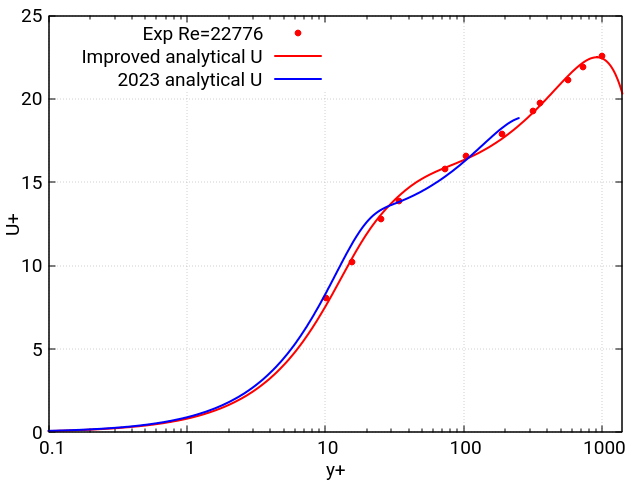}
 \caption{\label{fig:wei1} Improved analytical solution, Eq. (\ref{eq:u_sol2})  (red line), original Fedoseyev( 2023) analytical solution (blue line), and Wei  (1989) experiment (red dots) comparison for  Re = 22776. The comparison is for  the full experimental range of $y^{+}$=1400 for improved solution, while the original  solution is in the range up to $y^{+}$=300.
 The improved analytical solution (red line) passes closely to experimental data points (red dots).}
\end{center}
\vspace {-8mm}
\end{figure}
\begin{figure}[ht]
\begin{center}
\includegraphics[width=0.60\linewidth]{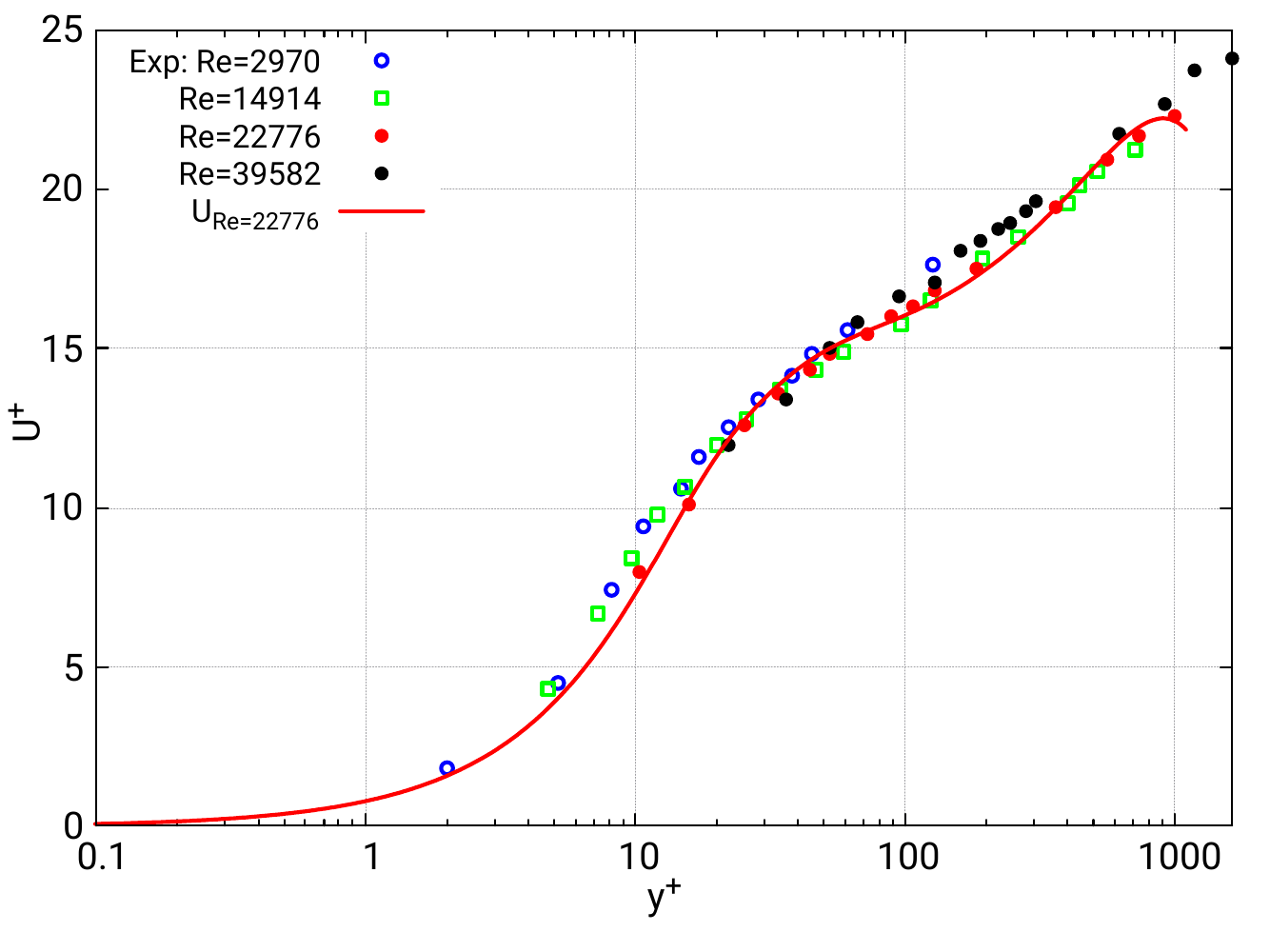}
\end{center}
\caption{\label{fig:wei2}Mean velocity profiles in turbulent boundary layer from Wei (1989) experiment, non-dimensionalized on inner variables,
for Reynolds numbers Re = 2970 (circles), 14914 (squares), 22776 (red dots), 39582 (black dots), and improved analytical solution of AHE, Eq. (\ref{eq:u_sol2}), for $Re=22776$ (red line). } 
\end{figure}

%
%

\noindent {\bf Comparison with Wei (1989) experimental data}. As an example, the analytical solution $U$ (blue line) for  Re = 22776 from Fedoseyev (2023) and the experimental data for mean velocity, Wei (1989) (red dots), are shown in Figure \ref{fig:wei1}. The comparisons are provided in  $(U^+, y^+)$ coordinates.  
The parameter $y^{+}$ is  $y^{+}= {yu_{\tau}}/{\nu}$, where $u_{\tau}$ is so called
friction velocity, y is the absolute distance from the wall, and $\nu$
is the kinematic viscosity. One can interpret $y^+$ as a local Reynolds
number. The friction velocity $u_{\tau}$ is defined as 
\begin{equation}\label{eq:utau}
u_{\tau}=\sqrt{ {\tau_{w}}/{\rho}},
\end{equation}

\noindent where wall shear stress $\tau_w$, $\rho$ is fluid density, $\tau_{w}=\rho\nu\frac{dU}{dy}$
at y=0, and the dimensionless velocity $U^{+}$ is given by $U^{+}= {u}/{u_{\tau}}$. 
The analytical solution $U$ (blue line), Eq.(\ref{eq:u_sol}), missed some of the experimental data points (red dots), and needs an improvement.
%
%
\section {Improved Analytical Solution Formula}

In the analytical solution Eq.(\ref{eq:u_sol}), Fedoseyev (2023), the sign of  $\delta = {\sqrt{\tau^{*}\nu}}/{L_0}$ was assumed to be positive. However, if  we take $\delta$  with a negative sign, we obtain another solution, $U^i_{T}=c_1e^{-e^{-y/\delta}}+c_2$. Combining this new term with the  original superexponential term, we arrive at an improved formula
\begin{equation}\label{eq:u_sol1}
U^i=U_{0}\left[\gamma\left(e^{1-e^{-y/\delta_1}}-e^{1-e^{y/\delta_1}}\right)+(1-\gamma)4y(L-y)/L^{2}\right], 
\end{equation}
where $\delta_1$ is different than $\delta$ in Eq.(\ref{eq:u_sol}), $\delta_1= 2 \delta /e$, to keep the same derivative $U^i_y$ value at $y=0$.
Substituting $\delta_1$ into Eq.(\ref{eq:u_sol2}), we obtain the  formula

\begin{equation}\label{eq:u_sol2}
U^i=U_{0}\left[\gamma\left(e^{1-e^{-ey/2\delta}} -e^{1-e^{ey/2\delta}}\right)+ (1-\gamma)4y(L-y)/L^{2}\right]. 
\end{equation}

%
%
\section {Comparison of Improved Analytical Solution with Experimental Data}
%
%

\subsection{Comparison of improved solution with Wei (1989) experimental data} 
The \cite{Wei_1989} experiments were done 
in turbulent channel flows over the range of Reynolds number
from 3000 to 40000 based on a channel half-width. The working fluid was distilled water.
The comparisons are provided in $(U^+, y^+)$ coordinates. 

The parameter $\gamma$ is calculated  through the minimization principle, Fedoseyev (2024), and is $\gamma=0.65$
for the experiment. The improved analytical solution $U^i$ (red line) for  Re = 22776 is compared with the experimental data (red dots), Figure \ref{fig:wei1} . A comparison for improved solution is extended to the full experimental range of $y^{+}=1400$, compared to \cite{Fedoseyev_2023}, where the range was $y^{+}=300$.
The improved analytical solution  (red line), given by Eq.(\ref{eq:u_sol2}), compares better with experimental data (red points), than previous solution (blue line), given by Eq.(\ref{eq:u_sol}). The improved analytical solution (red line) passes closely to the experimental data points (red dots).

\begin{figure}[h!]
	\begin{center}
	\includegraphics[width=0.49\linewidth]{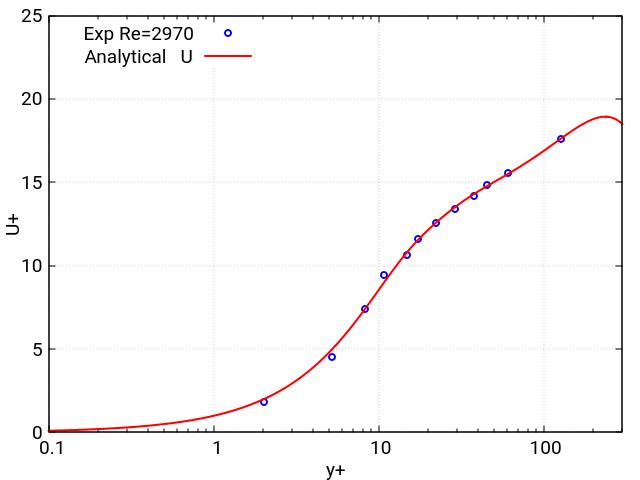}
        \includegraphics[width=0.49\linewidth]{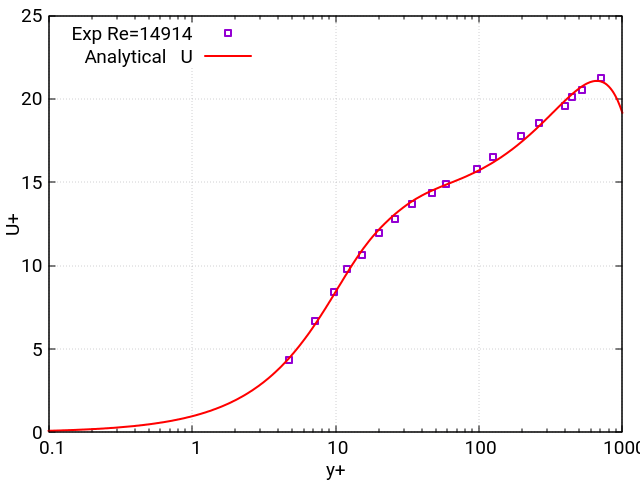}
        \includegraphics[width=0.49\linewidth]{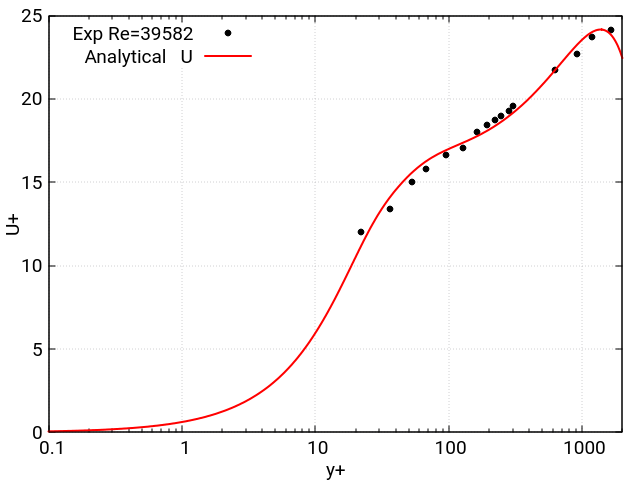}
	\caption{\label{fig:wei3} Improved analytical solutions (red line) compared with  Wei (1989) experiments for  Re = 2970 (circles); 14914
(squares) and 39582 (black dots). The comparison is for the full experimental range of $y^{+}$ for each experiment.}
	\end{center}
\end{figure}

Comparisons of the improved analytical solution for  Re = 22776 with other three Wei (1989) experiments are shown in Figure \ref{fig:wei2} .  Comparisons with each of separate Wei (1989) experiments for different Re numbers are shown in Figure \ref{fig:wei3} .
The analytical solution describes a complete flow velocity for every experimental Reynolds number, including the inner boundary layer (BL) region (viscous sublayer),  and near-middle (buffer) BL region, far-middle (inner) BL region, and outer (non-linear, essentially inviscid) region. 
The analytical solution demonstrates good agreement with experimental data.

%
%

\subsection{Comparison of improved solution with Nikuradse (1932) pipe flow experiment} 
Nikuradse's experiments were conducted at the Prandtl Laboratory to study turbulent flow in smooth circular pipes, covering Reynolds numbers from Re=27,000 to Re=970,000, \cite{Nikuradse_1932}. The working fluid was water.

The initial comparison with the analytical solution \cite{Fedoseyev_2023} shows significant deviation from experimental data near the end of the boundary layer at high Reynolds number. The parameters of the analytical solution were the following: $\gamma=0.71$ and $\delta=0.010$.

\begin{figure}[h]
	\begin{center}
	\includegraphics[width=0.60\linewidth]{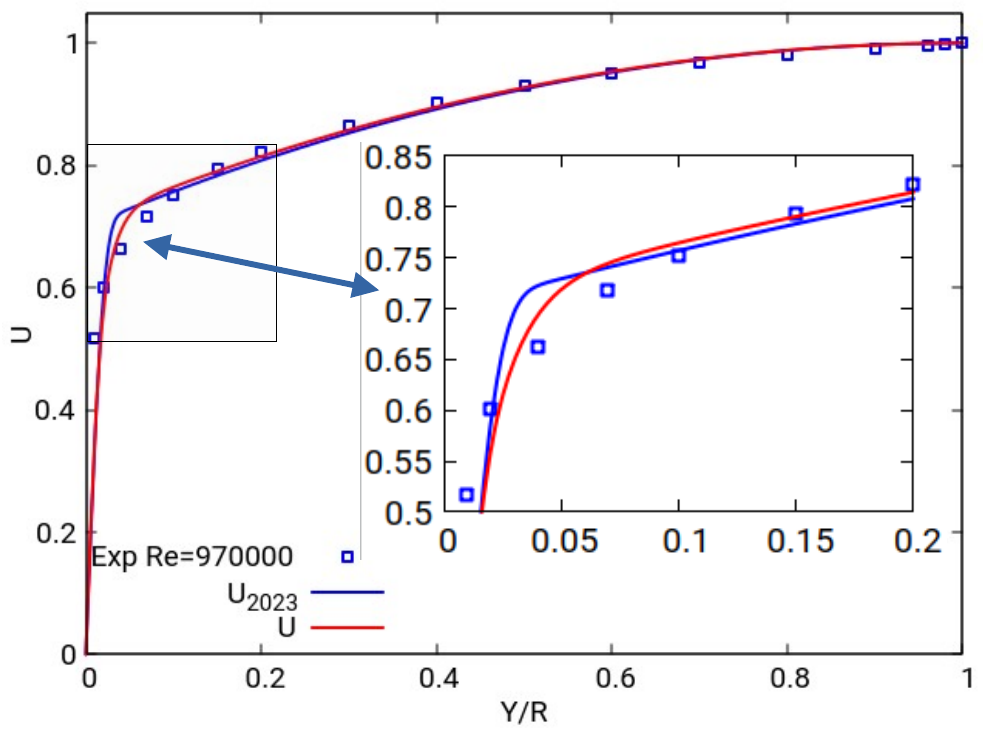}
	\caption{\label{fig:nik1}  Experimental data for mean velocity profiles from Nikuradse (1932) for Re=970,000 compared to improved analytical solution (red line), and the original Fedoseyev (2023) solution (blue line). The insert shows the area of improvement. Y/R is the ratio of distance from the wall to pipe radius.}
	\end{center}
\vspace {-5mm}
\end{figure}

The improved solution is in better agreement for experiments at higher values of Reynolds number, up to 
 $Re=10^6$, a range of nearly three orders of magnitude, compared to Wei (1989). Nikuradse (1932) experiments at Prandtl group have been done for Re=27,000 to Re=970,000. The comparison of Eq.(\ref{eq:u_sol2}) (improved) with Eq.(\ref{eq:u_sol}) (original) solutions show an improved agreement with experimental data for larger Reynolds numbers, Figure \ref{fig:nik1}.
  As illustrated in Figure~\ref{fig:nik1} , the discrepancy is reduced by a factor of two.
%
%
\subsection{Pipe Flow Experiment by Van Doorne at Re=7200}
In van Doorne experiment, a circular pipe with an inner diameter of 40 mm and a total length of 28 m was used. The working fluid was water. Due to a well designed contraction and thermal isolation of the pipe, the flow can be kept laminar up to Re = 60,000. All measurements were carried out at the distance of 26 m from the inlet, and stereoscopic-PIV was used \cite{Doorne_2007}. 

The material timescale $\tau^*$ for tap water is $\tau^* \approx  0.80$ s, as reported in  \cite{Fedoseyev_2024}.
The analytical solution can be applied to the pipe flow, where $\delta=0.047 \ll 1$. 

\begin{figure}[ht!]
\begin{center}
\includegraphics[width=0.60\textwidth]{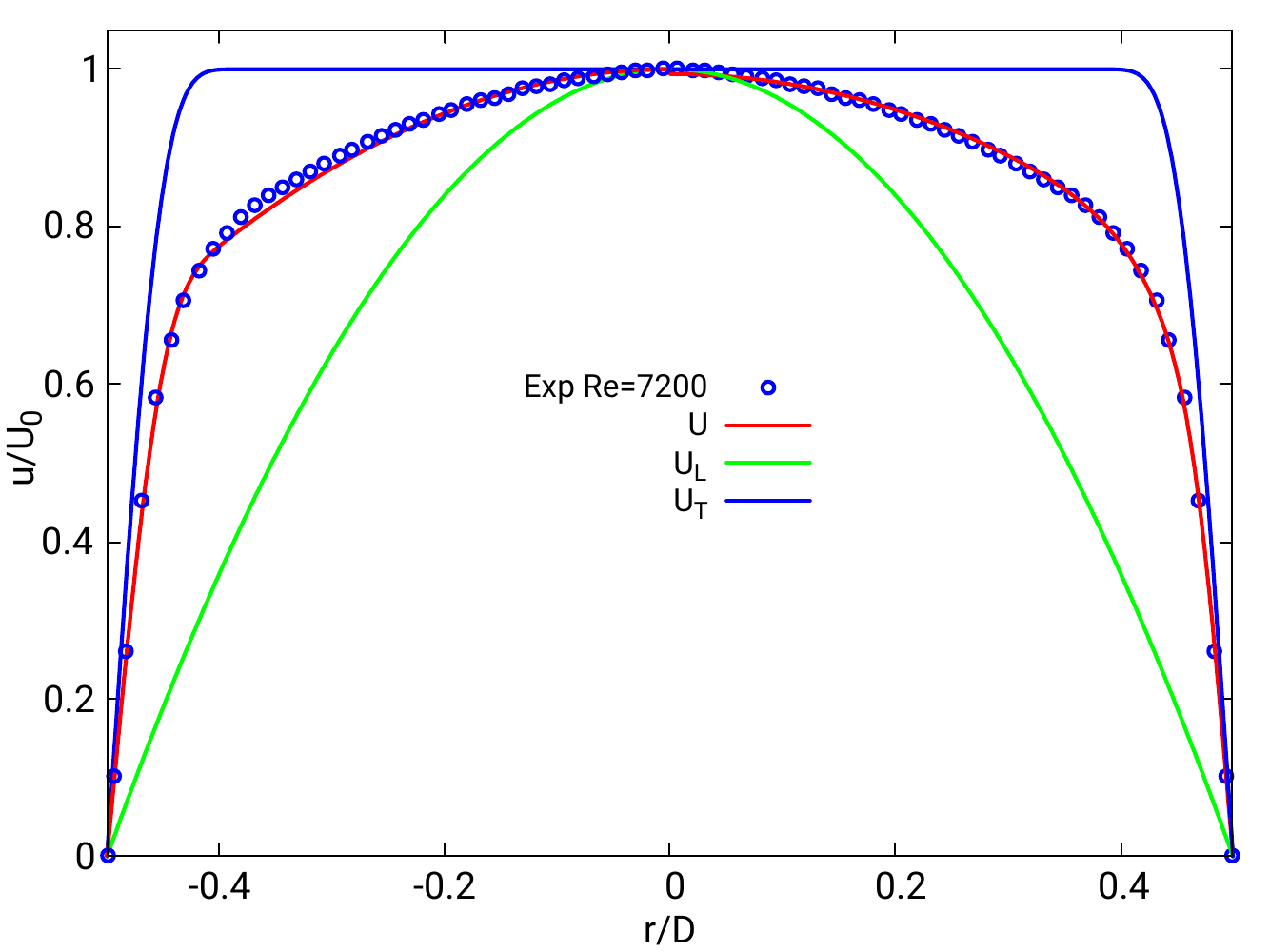}
\end{center}
\caption{\label{fig:doorne}
 Comparison of experimental data for streamwise velocity $U=u/V_0$ versus radius $r/D$  ($D$-diameter)  \cite{Doorne_2007}, (blue circles) at Re=7200, with the analytical solution (red line). The  left part of the $U$ plot is from \cite{Fedoseyev_2024}, and the right part is an improved formula, given by Eq.(\ref{eq:u_sol2}). Both  analytical solutions fit the experimental velocity profile well. The laminar (parabolic) solution (green line) and the turbulent (superexponential) solution (blue line) are also shown.  }
\end{figure}

Figure \ref{fig:doorne} shows the experimental data digitized from \cite{Doorne_2007}, along with several plots: (i) laminar (parabolic) flow profile (green line), (ii) turbulent (superexponential) flow profile (blue line) 
and (iii) AHE analytical solution (red line). The left part of the AHE plot is for $\gamma=0.68$ (minimal dissipation), and the right part is for $\gamma=0.65$ from \cite{Fedoseyev_2023}. The figure demonstrates that neither
the laminar nor turbulent solution fit the data, but the superposition $U$, Eq.(\ref{eq:u_sol2}),  provides good comparison to the experimental data.
\begin{figure}[ht!]
\begin{center}
\includegraphics[width=0.60\textwidth]{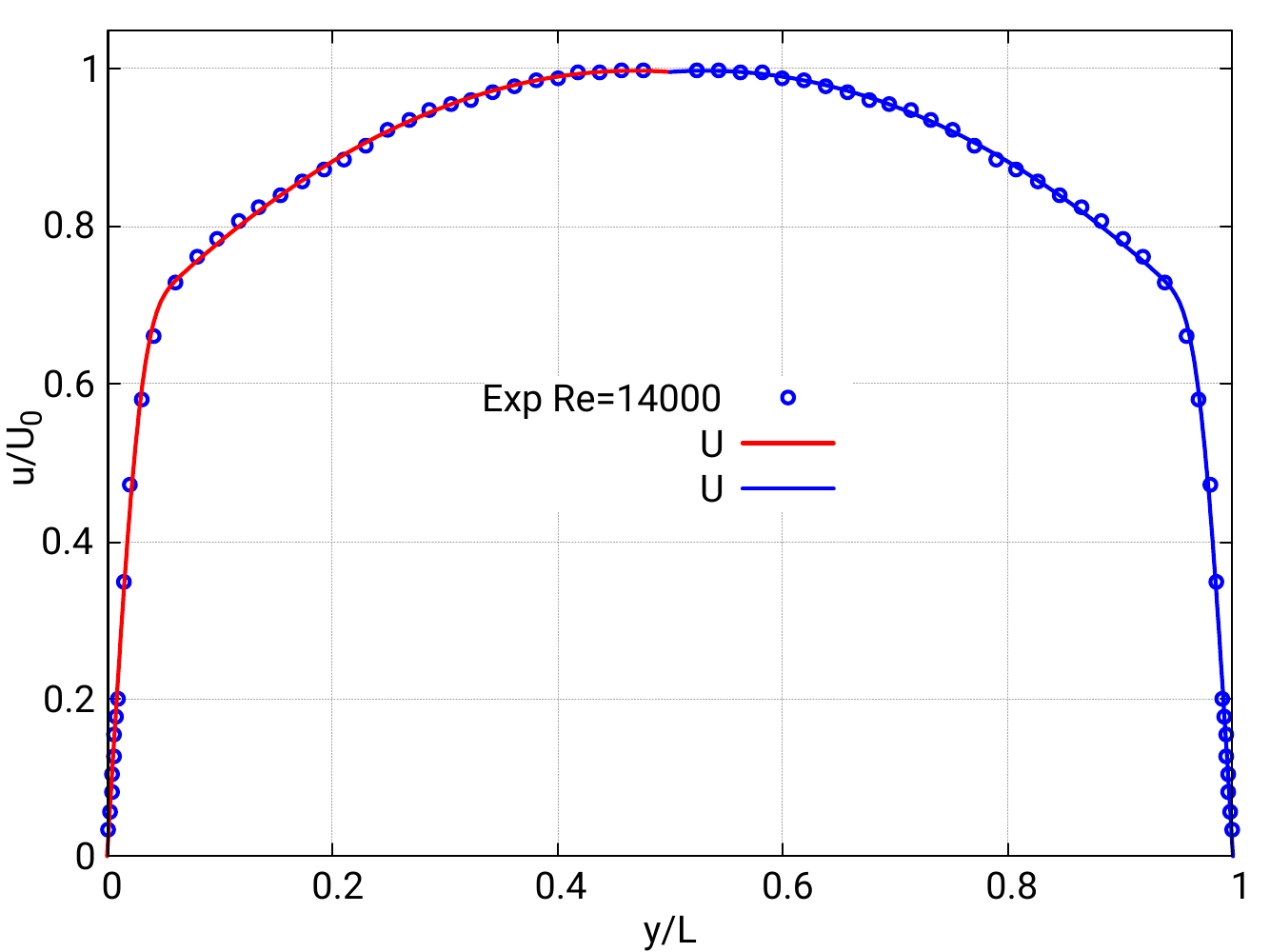}
\end{center}
\caption{\label{fig:pasch}
Comparison of streamwise velocity in a turbulent channel experiment \cite{Pasch_2023}, $Re=14000$ (blue circles), with the analytical solution $U$. The left portion of the $U$ curve (red line) follows \cite{Fedoseyev_2024}, while the right portion (blue line) uses the improved formula given by Eq.~(\ref{eq:u_sol2}). Both analytical solutions show good agreement with the experimental velocity profile.}
\end{figure}

%
%
\subsection{Channel Flow Experiment by Pasch at Re=15000}
A channel flow facility was used in the experiment. The test-section had a cross section of 300 mm x 25.2 mm (width x height) corresponding aspect ratio of 12 and an overall length of 4000 mm to ensure that the turbulent flow achieves a fully developed state.
The Laser Doppler Velocimetry (LDV) measurement concept used two
laser-beam pairs with different wavelengths cross building overlapping interference fringe systems in the same plane, to improve the accuracy. The tracer particles of a
nominal diameter of $d = 1 \mu m$. The experiments were performed at a Reynolds number of 7000 based on the centerline velocity of $U_{cl} = 8.4 m/s$
and the channel half height. This corresponds to $Re=14000$ for the channel height.

Figure \ref{fig:pasch} shows the streamwise velocity in a turbulent channel experiment by \cite{Pasch_2023} at $Re=14000$, along with the  analytical solution $U$, given by Eq.(\ref{eq:u_sol2}). The coefficient $\gamma=0.62$ was obtained by the minimum of  viscous dissipation  principle \cite{Fedoseyev_2024}. The parameter $\delta$ is $\delta=0.033$, the working fluid was air. The analytical solution shows good agreement with experimental data.

%
%
\subsection{Turbulent Pipe Flow Experiment by Zagarola (1996)}
An experimental investigation was conducted using the Princeton Superpipe setup, \cite{Zagarola_1996}, to determine the scaling of the mean velocity profile in fully developed turbulent pipe flow.
The working fluid was air. The circular pipe diameter was 12.9 cm. The flow was assumed to be incompressible since the maximum Mach number was less
than 0.08 for all surveys. The temperature was 
near ambient (295-300 K) and the pressure was varied between 1 and 186 atm.

\begin{figure}[ht!]
\begin{center}
\includegraphics[width=0.49\textwidth]{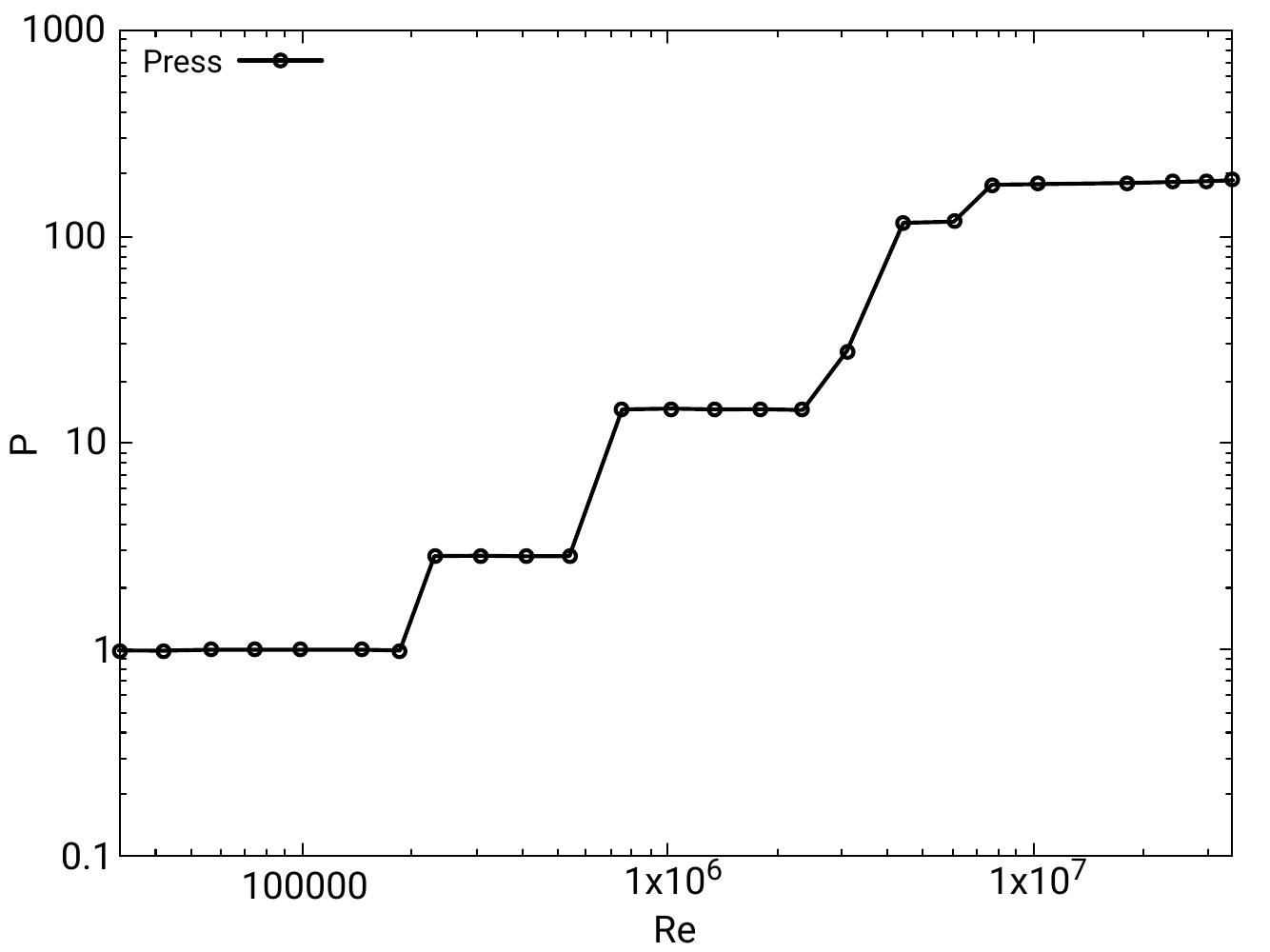}
\includegraphics[width=0.49\textwidth]{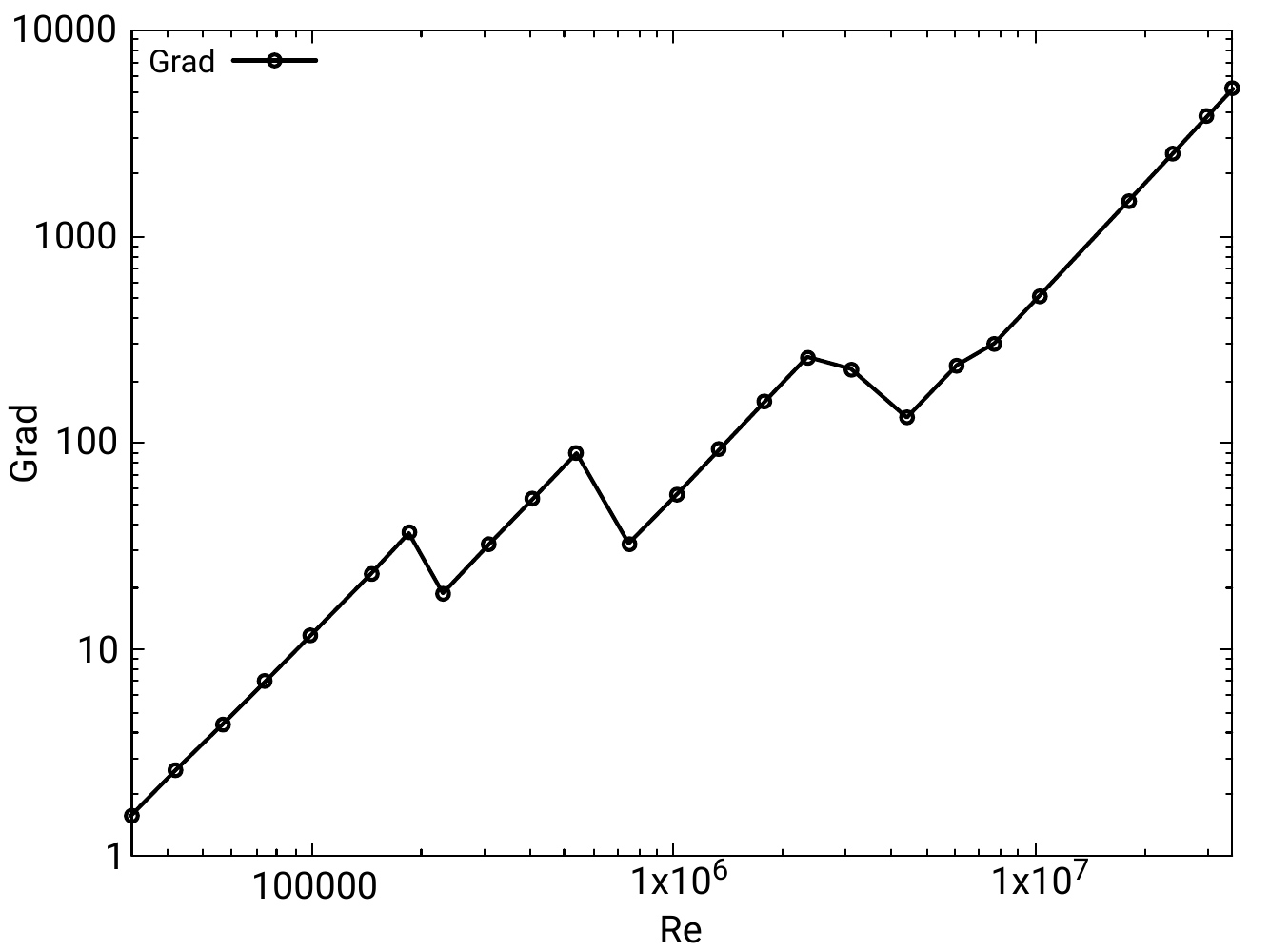}
\hspace{4cm} (a) \hspace{7cm} (b) \hspace{5cm} 
\includegraphics[width=0.49\textwidth]{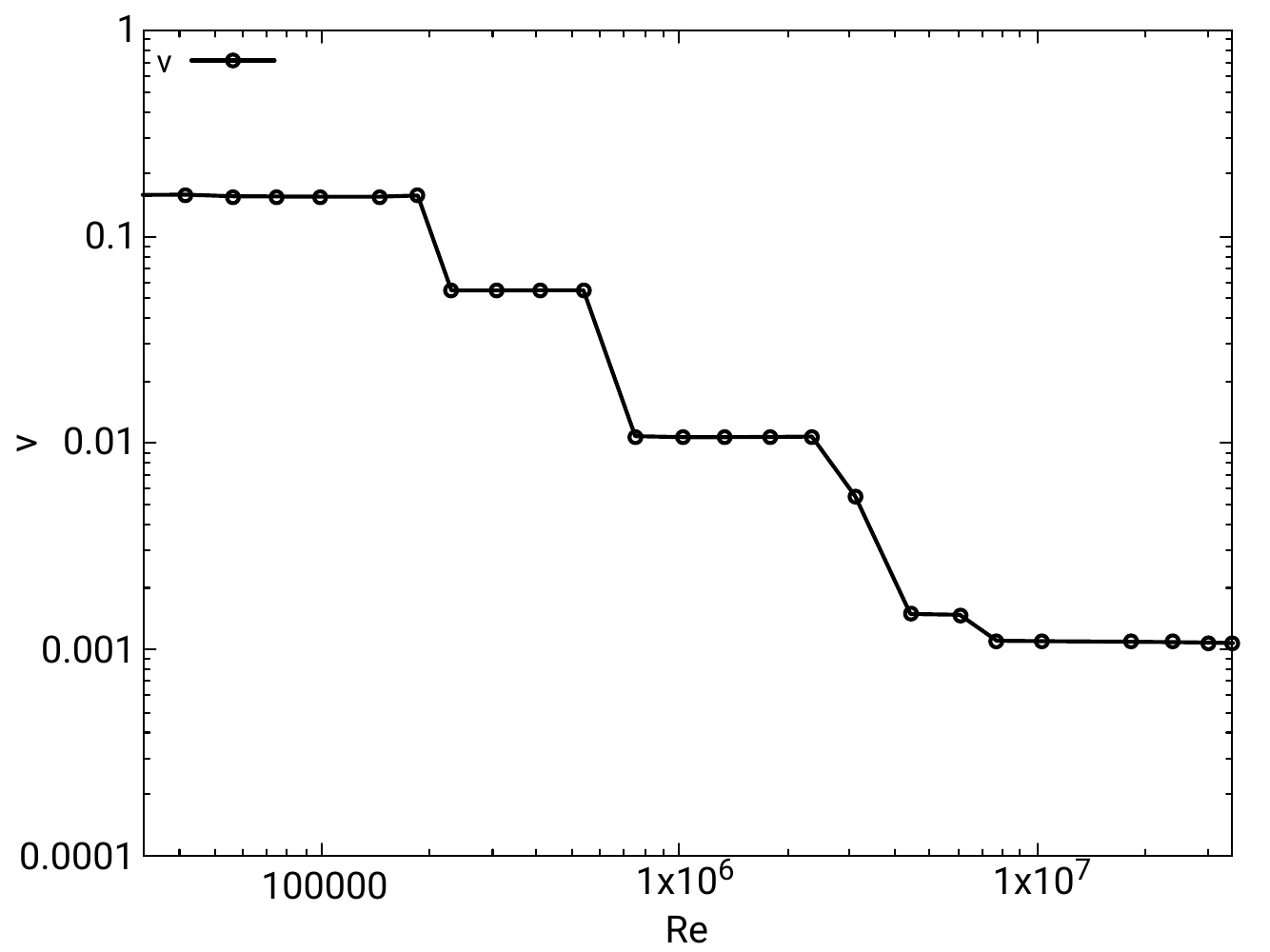}
\includegraphics[width=0.49\textwidth]{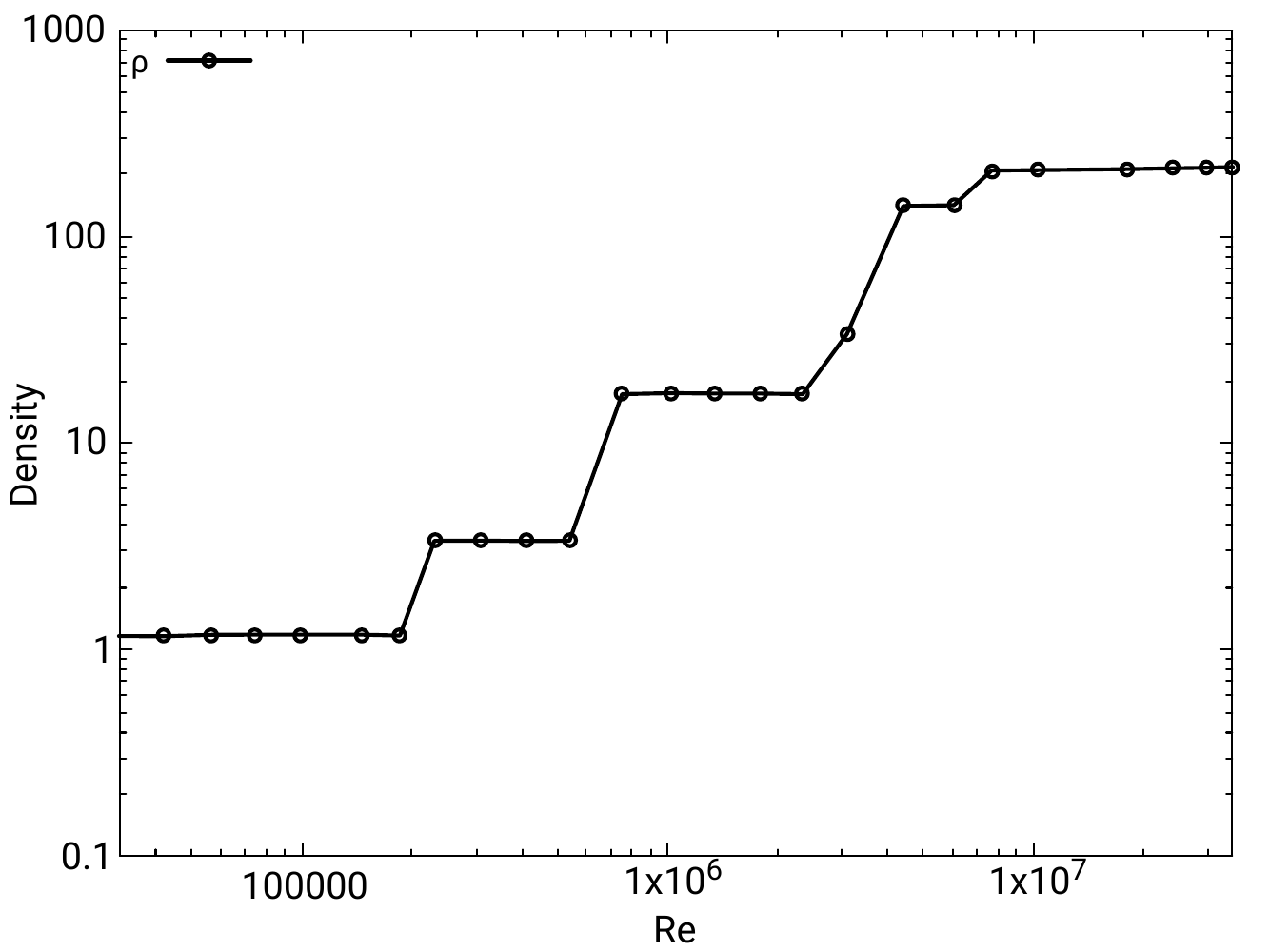}
\hspace{4cm} (c) \hspace{7cm} (d) \hspace{5cm} 
\includegraphics[width=0.49\textwidth]{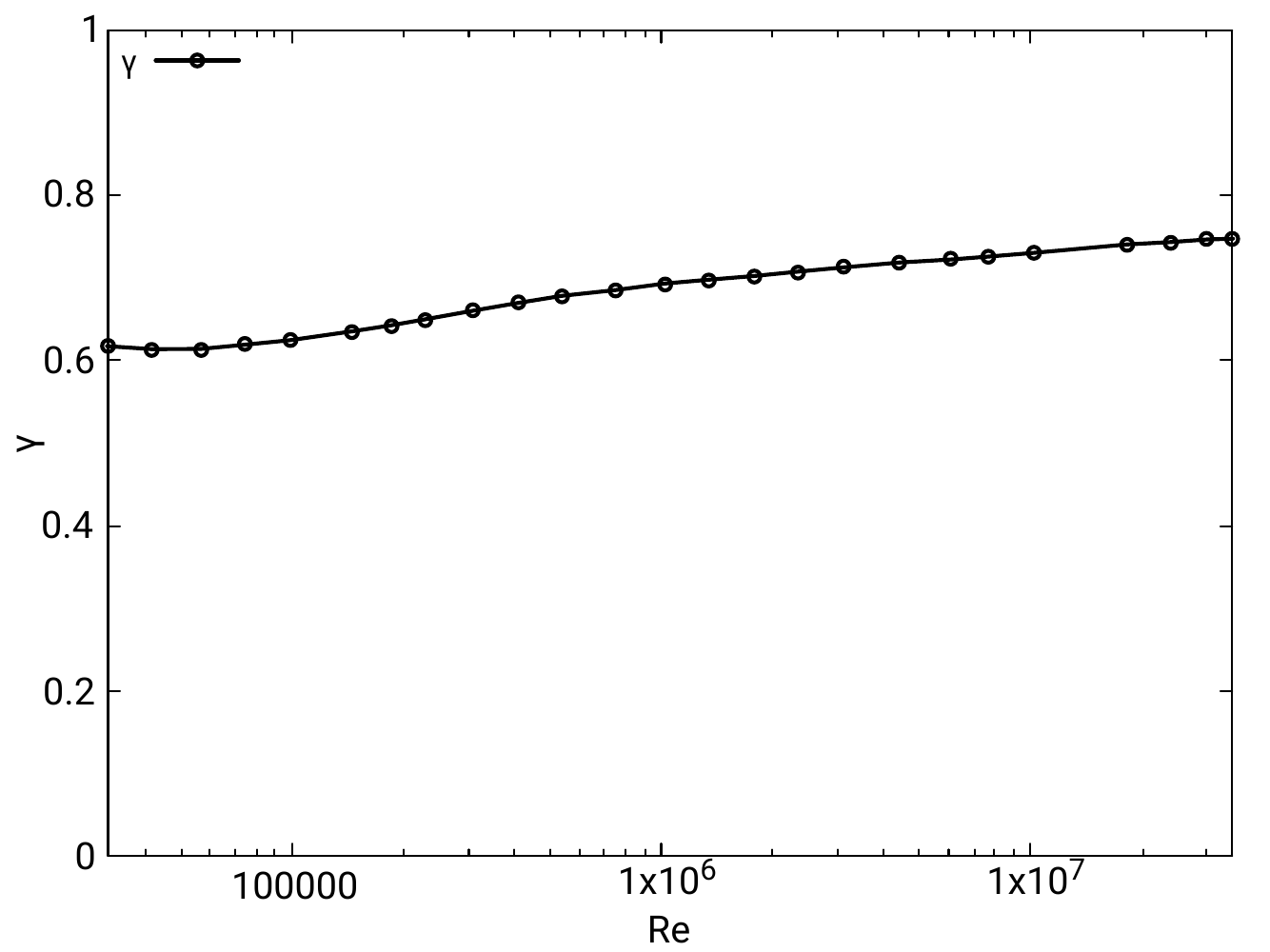}
\includegraphics[width=0.49\textwidth]{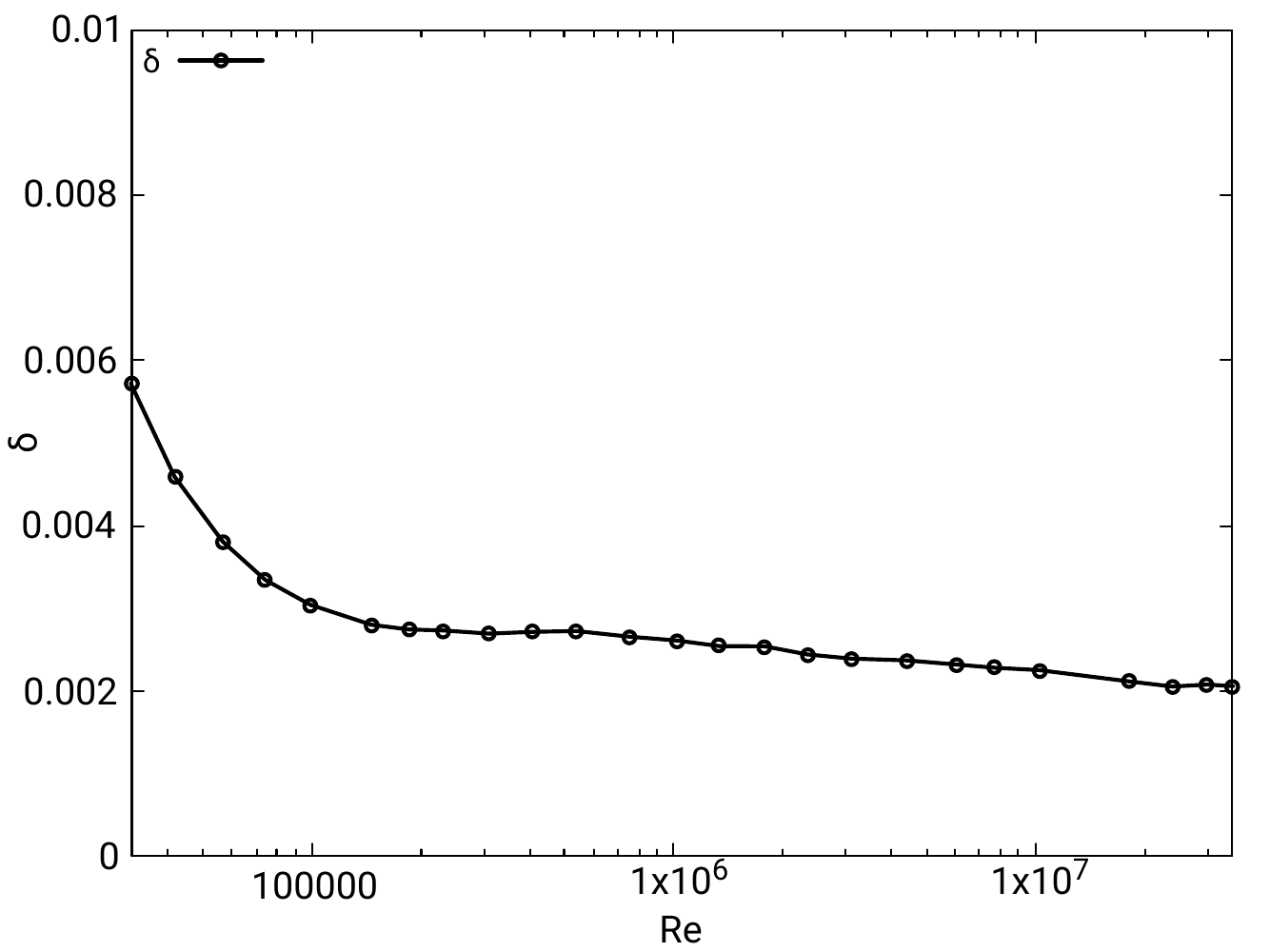}
\hspace{4cm} (e) \hspace{7cm} (f) \hspace{5cm} 
\end{center}
\caption{\label{fig:superexp}
Parameters for the Princeton Superpipe turbulent flow experiments \cite{Zagarola_1996} (26 experiments spanning $Re=3.15\cdot 10^4$ to $Re=3.52\cdot10^7$) vary by several orders of magnitude: (a) pressure [atm], (b) pressure gradient [Pa/m], (c) kinematic viscosity [cm$^2$/s], and (d) density [kg/m$^3$]. These figures do not show a dependence of the parameter on $Re$, but instead present its value for each individual experiment. Note that the analytical solution parameters $\gamma$ (e) and $\delta$ (f) do not vary significantly between experiments.  }
\end{figure}

Measurements of the mean velocity profiles and static pressure gradients were performed at 26 different Reynolds numbers, ranging from $3.15\cdot 10^4$ to $3.52\cdot10^7$. The experimental parameters vary by several orders of magnitude, Figure~\ref{fig:superexp}(a-d). Nevertheless, the analytical solution parameters $\gamma$, and $\delta$ do not vary so significantly, Figure~\ref{fig:superexp}(e-f).

\begin{figure}[ht!]
\begin{center}
\includegraphics[width=0.70\textwidth]{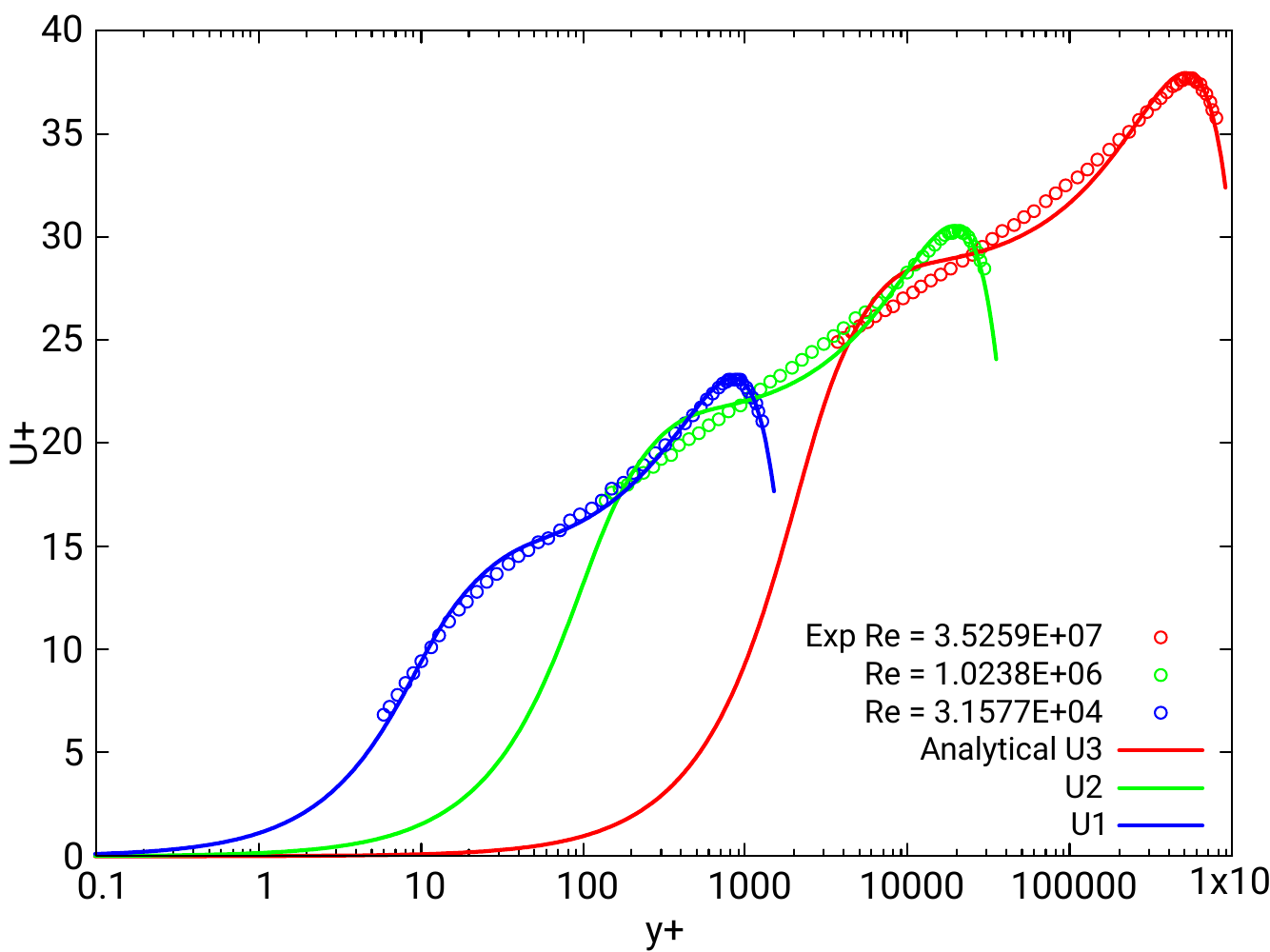}
\end{center}
\caption{\label{fig:super}
Comparison of the streamwise velocity in turbulent pipe flow from the Princeton Superpipe experiment \cite{Zagarola_1996}, $Re=3.15\cdot 10^4$ (blue circles), $Re=1.02\cdot 10^6$ (green circles), and $Re=3.52\cdot10^7$ (red circles). The improved analytical solution $U$ is shown for each case in the corresponding color, with a maximum error of 4\% observed at the highest Reynolds number.}
\end{figure}

We have constructed the analytical solution for all 26 experiments and present examples for the lowest, highest, and one intermediate Reynolds numbers. Figure \ref{fig:super} shows the streamwise velocity in a turbulent channel experiment for minimal Reynolds number $Re=3.15\cdot 10^4$, the medium $Re=1.02\cdot 10^6$, and the maximum $Re=3.52\cdot10^7$, along with the improved analytical solution $U$. The coefficient $\gamma$ varied between between 0.62 and 0.75, for all the cases, and the parameter $\delta$ varied between 0.0058 and 0.0027. 
The analytical solution shows agreement with the experimental data within 4\% observed at the highest Reynolds number.

%
%

\section {Discussion}
This work shows an improvement of the analytical solution and demonstrates better agreement with experimental data compared to the original approach, \cite{Fedoseyev_2023}.

%
%

\subsection{The coefficient $\gamma$\label{sec:gamma}}
The superposition parameter $\gamma$ is calculated using the proposed principle of minimal  viscous dissipation of energy 
relative to the dissipation value on the wall \cite{Fedoseyev_2024}. 
The minimum of integral gives a solution for $\gamma$: 
\begin{equation}\label{eq:diss_tot}
\varepsilon_{T} =  \frac{ 1}{U_{y}(0)^2} \int^L_0 U_y^2 dy \cdot
\frac{ 1}{L} \int^L_0 U\ dy,
\end{equation}
 where $U(y)$ is presented by Eq.(\ref{eq:u_sol2}). The parameter $\gamma$ is within the interval 0.6 to 0.7 for most of the cases considered.

%
%
\subsection{The similarity parameter $\delta$\label{sec:delta}}

The second similarity parameter $\delta$ does not depend on the Reynolds number, as shown in the data by Wei and Willmarth experiment \cite{Wei_1989}. The experimental data for different Reynolds numbers collapse closely onto the same curve, Figure \ref{fig:wei2} . Knowing $\delta$, one can determine the timescale parameter $\tau^*$, which is a material property.

%
%
\subsection{The timescale parameter $\tau^*$ \label{sec:tau}}

\begin{figure} 
\begin{center}
\includegraphics[width=0.60\textwidth]{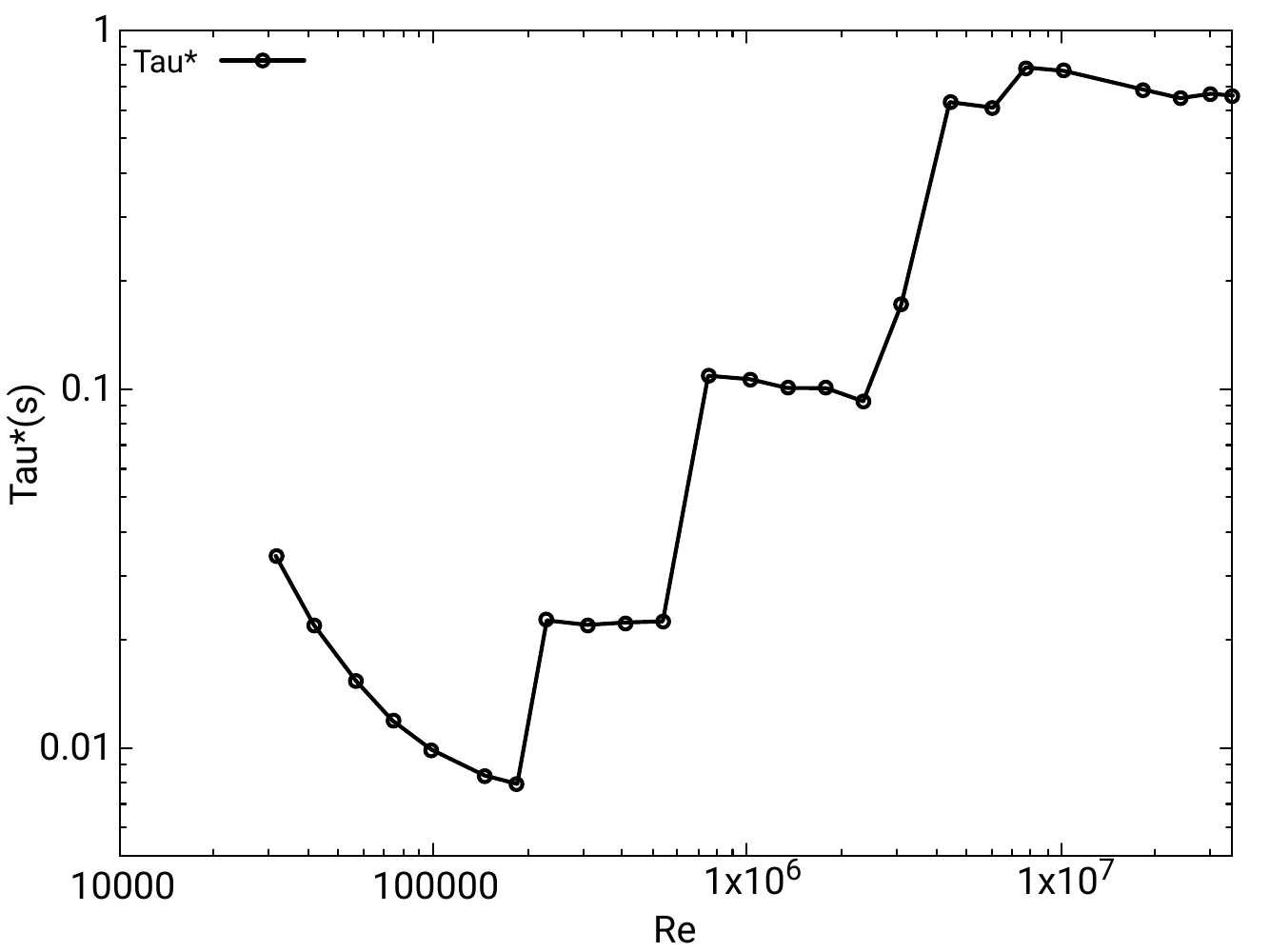}
\end{center}
\caption{\label{fig:tau}
The timescale parameter $\tau^*$ was calculated for all 26 Superpipe experiments at different Reynolds numbers. While the value of $\tau^*$ varies significantly, it defines the parameter $\delta = \sqrt{\tau^* \nu}/L_0$, which shows little variation across all experiments (Figure~\ref{fig:superexp}(f)). It should be emphasized that this is not a dependence of $\tau^*$ on $Re$ since $\tau^*$ is a material property and does not depend on $Re$, but rather the value of $\tau^*$ for each individual experiment. Similarly, Figure~\ref{fig:superexp}(a) does not show a dependence of pressure $P$ on $Re$, but instead presents the value of $P$ for each individual experiment.
}
\end{figure}

By analyzing several experiments and simulations \cite{Wei_1989}, \cite{Doorne_2007}, \cite{Nikuradse_1932}, \cite{Fedoseyev_2023}, we determined that the dimensional timescale  $\tau^* = \delta^2 L_0^2 / \nu$   is $\tau^* = 0.40 \pm 0.05$ s for distilled water, and approximately $\tau^* \approx  0.80$ s for tap water \cite{Fedoseyev_2024}. 
The estimation of $\tau^*$ for air was obtained from \cite{Pasch_2023} for  $\delta = 0.033$. Assuming viscosity and pressure corresponding to standard atmospheric conditions at 20\textdegree C, we obtain $\tau^* = 0.047$ s. While the value of $\tau^*$ varies significantly across the 26 experiments of \cite{Zagarola_1996} due to differences in pressure, and respectively the density and viscosity, Figure~\ref{fig:tau}, the more relevant parameter appears to be $l=\sqrt{\tau^* \nu}$, which defines $\delta$ and varies little among all experiments (Figure~\ref{fig:superexp}(f)). This parameter $l=\sqrt{\tau^* \nu}$ also can considered as a material property, together with $\tau^*$. 
%
%
\subsection{Turbulent Boundary Layer \label{sec:TBL}}
\begin{figure} 
\begin{center}
\includegraphics[width=0.60\textwidth]{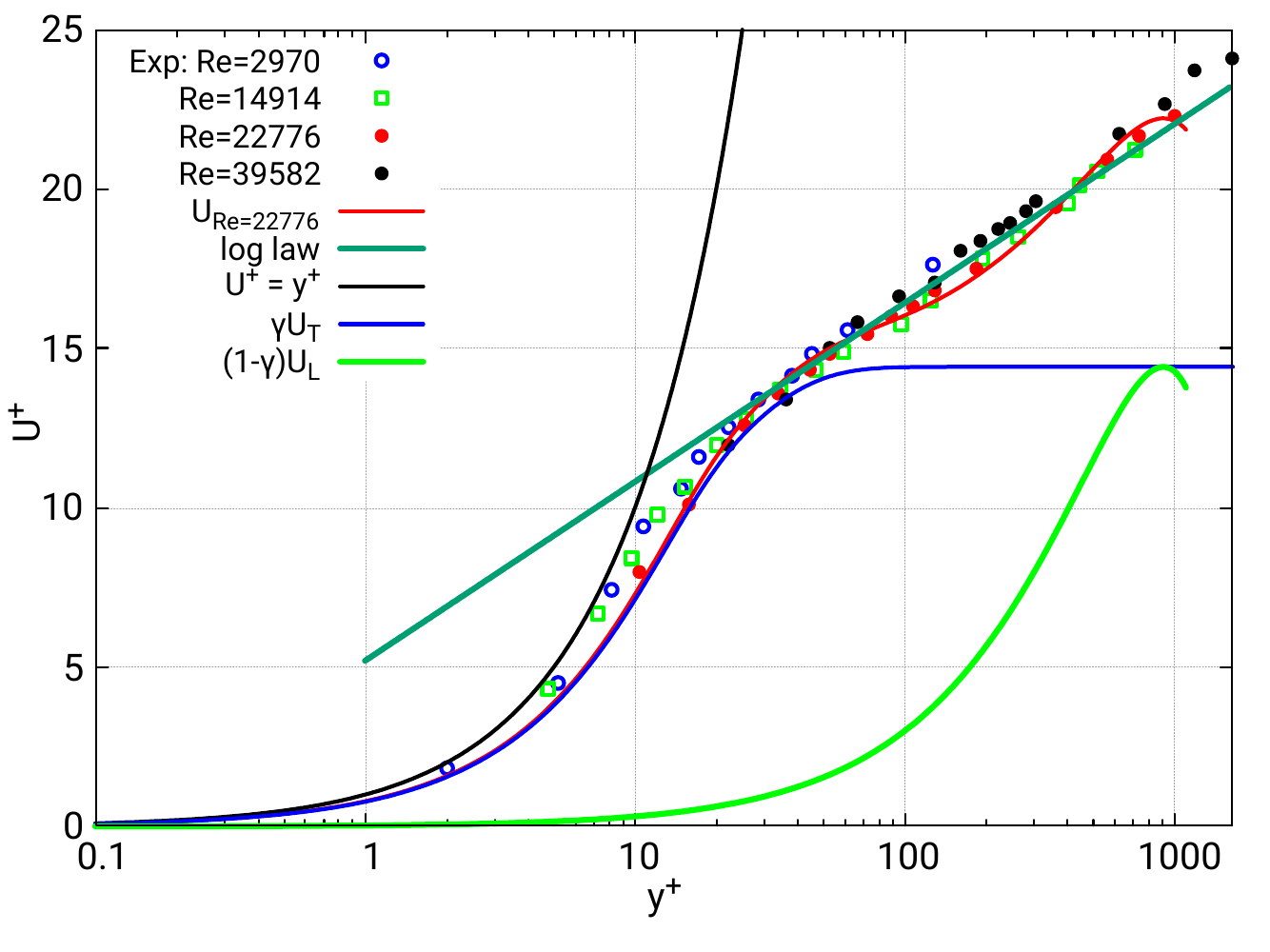}
\end{center}
\caption{\label{fig:law}
Velocity profiles for turbulent flow in channel: experimental data (dots) by \cite{Wei_1989}, improved analytical solution $U$ for Re=22776 (red line) and its constituents, the turbulent solution $\gamma U_{T}$ and the laminar solution $(1-\gamma)U_{L}$; log law by von Karman $U^+$ = 1/k log $y^+$ + B, k = 0.41, B = 5.2 (log law line, dark green), and linear law $U^+ = y^+$ (black line). }
\end{figure}

One can see (Figure~\ref{fig:law}, red line) that the analytical solution of the AHE accurately represents the turbulent boundary layer across all regions. The linear law holds in the range $0 < y^+ < 5$, where the parabolic profile $(1-\gamma)U_{L}$ (the laminar solution) is negligible, and the analytical solution for small $y/\delta$ reduces to
\begin{eqnarray*}
U =  \gamma U_{T}  = U_{0}\gamma y/\delta , 
\end{eqnarray*}
that is a linear law.
The near-middle (buffer) boundary layer region, $5 < y^+ < 30$, is strictly nonlinear, and the analytical solution agrees well with the experimental data.

In the far-middle (inner) boundary layer region, $30 < y^+ < 200$, the superexponential component $\gamma U_{T}$ approaches a nearly constant value of $\gamma U_0 \approx 14.4$ (Figure~\ref{fig:law}, blue line). Meanwhile, the AHE analytical solution (red line) exhibits variation due to the increasing contribution of the laminar component $(1-\gamma)U_{L}$ (Figure~\ref{fig:law}, black line). In this region the AHE solution is a parabolic function 
\begin{eqnarray*}
U = 14.4 + (1-\gamma)U_{L}= 14.4 + (1-\gamma)U_0 4y(L-y)/L^{2} = A + By + Cy^2,
\end{eqnarray*}
where $A=\gamma U_0 \approx 14.4$, $B= 4 (1-\gamma)U_0/L$, $C=-4 (1-\gamma)U_0/L^2$, and provides a closer fit to the experimental data (red dots) compared to the classical logarithmic von Karman law (dark green line).

The outer region (nonlinear and essentially inviscid) begins at $y^+ > 200$ and extends to the centerline, where the analytical solution also shows good agreement with the experimental data.

%
%
\section*{Conclusions} 

The improved analytical solution for turbulent channel and circular pipe flow has been presented and validated against multiple experiments, showing improved agreement with mean velocity data. It accurately captures the velocity behavior across the entire turbulent boundary layer and into the external flow, from the inner viscous sublayer to the outer region. 

Several additional comparisons with experimental data were carried out, particularly using air as the working fluid in the Princeton Superpipe experiments \cite{Zagarola_1996}, covering Reynolds numbers from $Re = 3.2 \cdot 10^4$ to $Re = 3.5 \cdot 10^7$.
The improved formula enables the accurate description of the turbulent channel or circular pipe boundary layer for much higher Reynolds numbers, up to $3.5 \cdot 10^7$, with an error below 4\% at the upper end.

%
%
\bibliographystyle{jfm}

\end{document}